\documentclass{elsart}
\usepackage{latexsym,epsf,cite,psfig,amssymb}

\begin{document}
\begin{frontmatter}

\title{Equation of state for systems with Goldstone bosons}
\author[aa]{Massimo Campostrini}
\author[bb]{Martin Hasenbusch}
\author[cc]{Andrea Pelissetto}
\author[aa]{Paolo Rossi}
\author[aa]{Ettore Vicari}

\address[aa]{Dipartimento di Fisica, Universit\`a di Pisa, and
              INFN, Sez. di Pisa, \\ I-56126 Pisa, Italy}
\address[bb]{NIC/DESY Zeuthen, Platanenallee 6,
              D-51738 Zeuthen, Germany}
\address[cc]{Dipartimento di Fisica, Universit\`a di Roma La Sapienza,
  \\ and INFN, Sez. di Roma I, I-00185 Roma, Italy} 

\begin{abstract}
We discuss some recent determinations of the equation of state for 
the XY and the Heisenberg universality class.
\end{abstract}

\begin{keyword}
Heisenberg model \sep XY model \sep equation of state \sep critical behavior 
 
\PACS 67.40.-w \sep 64.60.Fr \sep 11.15.Me \sep 05.70.Jk
\end{keyword}
\end{frontmatter}

\newcommand{\be}{\begin{equation}}
\newcommand{\ee}{\end{equation}}
\newcommand{\bea}{\begin{eqnarray}}
\newcommand{\eea}{\end{eqnarray}}
\newcommand{\<}{\langle}
\renewcommand{\>}{\rangle}

\def\spose#1{\hbox to 0pt{#1\hss}}
\def\ltapprox{\mathrel{\spose{\lower 3pt\hbox{$\mathchar"218$}}
 \raise 2.0pt\hbox{$\mathchar"13C$}}}
\def\gtapprox{\mathrel{\spose{\lower 3pt\hbox{$\mathchar"218$}}
 \raise 2.0pt\hbox{$\mathchar"13E$}}}

\def\bsigma{\mbox{\protect\boldmath $\sigma$}}
\def\btau{\mbox{\protect\boldmath $\tau$}}
\def\bphi{\mbox{\protect\boldmath $\phi$}}
\def\bz{\mbox{\protect\boldmath $z$}}
\def\bw{\mbox{\protect\boldmath $w$}}
\def\hatp{\hat p}
\def\hatl{\hat l}
\def\smfrac#1#2{{\textstyle\frac{#1}{#2}}}
\def\case#1#2{{\textstyle\frac{#1}{#2}}}

\def\msbar{ {\overline{\hbox{\scriptsize MS}}} }
\def\normalmsbar{ {\overline{\hbox{\normalsize MS}}} }

\newcommand{\R}{\hbox{{\rm I}\kern-.2em\hbox{\rm R}}}

\newcommand{\reff}[1]{(\ref{#1})}


In the last few years there has been a significant progress in the 
determination of the critical properties of $O(N)$ models; see, e.g., 
Ref.~\cite{review} for a comprehensive review. First of all, 
high-precision estimates of critical exponents and of several 
high-temperature universal ratios have been obtained by using 
{\em improved} Hamiltonians. Improved models are such that the 
leading nonanalytic correction is absent in the expansion of any 
thermodynamic quantity near the critical point. The idea is quite 
old \cite{CFN-82,GR-84,FC-85}. However, the early attempts 
that used high-temperature techniques were not able to 
determine improved models with high accuracy, so that 
final results did not significantly improve the estimates of 
standard analyses. Recently
\cite{BFMM-98,HPV-99,BFMMPR-99,Hasenbusch-99,HT-99,CHPRV-01,%
Hasenbusch-01,CHPRV-02}, it has been realized that Monte Carlo 
simulations using finite-size scaling techniques are very effective for 
this purpose, obtaining accurate determinations of several improved 
models in the Ising, XY, and O(3) universality class. 
Once an improved models is accurately determined, one can use 
standard high-temperature techniques in order to obtain very precise 
determinations of the critical exponents. For instance, for the 
experimentally relevant cases, we obtained 
\cite{CPRV-99,CPRV-00,CHPRV-01,CHPRV-02}:
\begin{eqnarray}
&& \gamma = 1.2373(2), \qquad\qquad \nu = 0.63012(16), 
       \qquad \qquad N = 1, 
\nonumber \\
&& \gamma = 1.3177(5), \qquad\qquad \nu = 0.67155(27), 
       \qquad \qquad N = 2,
\nonumber \\
&& \gamma = 1.3960(9), \qquad\qquad \nu = 0.7112(5), 
       \hphantom{12} \qquad \qquad N = 3.
\nonumber 
\end{eqnarray}
Beside the critical exponents, experiments may determine several 
other universal properties. We consider  here the equation of state 
that relates the magnetic field $\vec{H}$, the magnetization
$\vec{M}$, and the reduced temperature $t\equiv (T-T_c)/T_c$. 
In a neighborhood of the critical point $t=0$, $\vec{H}=0$, it can be written
in the scaling form
\be
\vec{H} = \left(B_c\right)^{-\delta} \vec{M} M^{\delta-1} f(x),
\qquad\qquad x \equiv t (M/B)^{-1/\beta},
\label{eqstfx}
\end{equation}
where $B_c$ and $B$ are the amplitudes of the magnetization on the critical
isotherm and on the coexistence curve,
\begin{eqnarray}
    M &=& B_c H^{1/\delta}\qquad\qquad t=0. \label{def-Bcconst} \\
    M &=& B (-t)^\beta\qquad\qquad  H=0,\,\, t<0. \label{def-Bconst}
\end{eqnarray}
With these choices, the coexistence line corresponds to $x=-1$, and 
$f(-1) = 0$, $f(0) = 1$. Alternatively, one can write 
\begin{equation}
\vec{H} = k_1 {\vec{M}\over M} |t|^{\beta\delta} F_{\pm} (|z|),
\qquad\qquad z \equiv k_2 M t^{-\beta},
\end{equation}
where $F_+(z)$ applies for $t>0$ and $F_-(|z|)$ for $t<0$. 
The constants $k_1$ and $k_2$ are fixed by requiring
\begin{equation} 
F_+(z) = z + {1\over6} z^3 + \sum_{n=3} {r_{2n}\over (2n-1)!} z^{2n-1}
\label{exp-Fpiu}
\end{equation}
for $z \to 0$ in the high-temperature phase.
The behavior of the functions $f(x)$ and $F_-(|z|)$ at the 
coexistence curve depends crucially on $N$.
For $N=1$ they vanish linearly. On the other hand, for $N\ge 2$, the presence 
of the Goldstone modes implies in three dimensions 
\cite{BW-73,BZ-76,WZ-75,SH-78,Lawrie-81}:
\be
f(x) \approx c_f (1 + x)^2.
\label{fx-coex}
\ee
The nature of the corrections to this behavior is not clear 
\cite{WZ-75,SH-78,Lawrie-81,PV-99}. In particular, logarithmic terms are 
expected \cite{PV-99}.

In order to obtain approximations of the equation of state, we parametrize the 
thermodynamic variables in terms of two parameters $\theta$ and $R$:
\begin{eqnarray}
M &=& m_0 R^\beta m(\theta) ,\qquad t = R(1-\theta^2), \qquad
H = h_0 R^{\beta\delta}h(\theta). \label{parrep}
\end{eqnarray}
Here, $m_0$ and $h_0$ are nonuniversal constants,
$m(\theta)$ and $h(\theta)$ are odd functions of $\theta$,
normalized so that
$m(\theta)=\theta+O(\theta^3)$ and $h(\theta)=\theta+O(\theta^3)$.
The variable $R$
is nonnegative and measures the distance from the critical point in
the $(t,H)$ plane, while the variable $\theta$ parametrizes the
displacement along the lines of constant $R$.
In particular, $\theta=0$ corresponds to the high-temperature 
line $t>0$, $H=0$, $\theta=1$ to the critical isotherm $t=0$, 
and $\theta = \theta_0$, where $\theta_0$ is the smallest positive zero of 
$h(\theta)$---it must satisfy of course $\theta_0 > 1$---to the 
coexistence line. Such a mapping has been extensively used in the Ising case
and provides accurate approximations if one uses low-order polynomials
for $m(\theta)$ and $h(\theta)$ 
\cite{Schofield-69,SLH-69,Josephson-69,GZ-97,CPRV-99,CHPV-01}. 
In systems with Goldstone bosons we must additionally ensure the 
condition (\ref{fx-coex}). For this purpose, it is enough to
require \cite{CPRV-00-2}
$h(\theta) \sim   \left( \theta_0 - \theta\right)^2$ for 
        $\theta \rightarrow \theta_0$.

In Refs. \cite{CPRV-00-2,CHPRV-01,CHPRV-02} we obtained the equation 
of state in the scaling limit by using two different approximation 
schemes for the functions $m(\theta)$ and $h(\theta)$:
\begin{eqnarray}
{\rm scheme}\quad({\rm A}):\qquad\qquad
&&m(\theta) = \theta
    \left(1 + \sum_{i=1}^n c_{i}\theta^{2i}\right), \nonumber \\
&&h(\theta) = \theta \left( 1 - \theta^2/\theta_0^2 \right)^2,
\label{scheme1}
\\[3mm]
{\rm scheme}\quad({\rm B}):\qquad\qquad
&&m(\theta) = \theta, \nonumber \\
&&h(\theta) = \theta
    \left(1 - \theta^2/\theta_0^2 \right)^2
    \left( 1 + \sum_{i=1}^n c_{i}\theta^{2i}\right).
\label{scheme2}
\end{eqnarray}

\begin{figure}
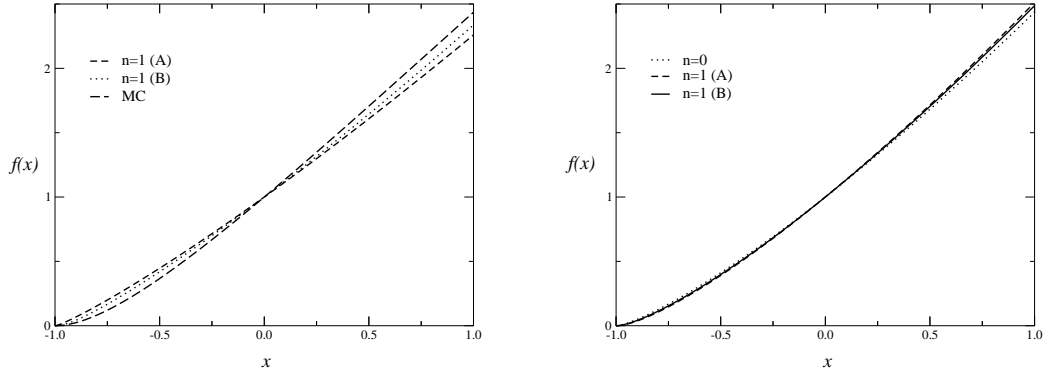

\begin{tabular}{cc}
 \hspace{-0.6cm} 
\psfig{figure=./fxXY.eps,angle=-0,width=0.46\linewidth}     &
 \hspace{0.5cm}
\psfig{figure=./fxH.eps,angle=-0,width=0.46\linewidth}
\end{tabular}
\caption{Graph of the function $f(x)$ for $N=2$ (left) and 
$N=3$ (right). For $N=2$ we also report the Monte Carlo results of 
Ref.~\cite{EHMS-00}.
}
\label{figura}
\end{figure}

\noindent The constants $c_i$ and $\theta_0$ were fixed by requiring
$F_+(z)$ to have the expansion (\ref{exp-Fpiu}), with the coefficients 
determined by high-temperature expansion techniques. Since we were 
able to compute accurately only $r_6$ and $r_8$, we used the two schemes
for $n=0$ and $n=1$. The results, especially those for $N=3$,
see Fig. \ref{figura}, are quite independent of the scheme used,
indicating the good convergence of the method.

By using the equation of state, one can determine several amplitude ratios. 
We mention here the experimentally relevant
\be
U_0 = {A^+\over A^-},
\qquad \qquad
R_\chi = {C^+ B^{\delta-1}\over B_c^\delta}, 
\nonumber 
\ee
where $C^+$ and $A^\pm$ are related to the critical behavior of 
the susceptibility $\chi$ and of the specific heat $C$ for $H=0$:
\begin{eqnarray}
&& \chi = C^+ t^{-\gamma},  \hskip 1.5truecm \qquad\qquad  t>0, \nonumber \\
&& C    = A^\pm (\pm t)^{-\alpha} + B  \qquad\qquad  \pm t>0. \nonumber
\end{eqnarray}
Using the approximate equation of state we obtain
\cite{CPRV-00,CHPRV-01,CHPRV-02}: 
$U_0 = 1.062(4)$, $R_\chi = 1.35(7)$ for  $N = 2$ and
$U_0 = 1.57(4)$, $R_\chi = 1.33(8)$  for $N = 3$.

For $N=3$ the approximate equation of state can be compared with 
experiments. We observe good qualitative and quantitative agreement.
For $N=2$ we can use our results to predict critical properties
of the $\lambda$-transition in ${}^4$He. In this case, the equation of 
state does not have a direct physical meaning, but we can still 
compare the predictions for the singular specific-heat ratio $U_0$. 
A precise determination of the exponent $\alpha$ and of $U_0$ 
was done recently by means of a calorimetric experiment in microgravity
\cite{LSNCI-96} (see also reference 4 in Ref. \cite{CHPRV-01})
obtaining $\alpha = -0.01056(38)$ and $U_0\approx 1.0442$.
The result for $U_0$ is lower than the theoretical one.
This is strictly related to the disagreement in the value of $\alpha$ 
(see also Ref. \cite{HES-01}).
Indeed, using hyperscaling we find $\alpha = -0.0146(8)$, that 
significantly differs from the experimental estimate.
The origin of this discrepancy is unclear and further theoretical and 
experimental investigations are needed. A new generation of 
experiments in microgravity environment that is currently in preparation
should clarify the issue on the experimental side \cite{Nissen-etal-98}.


\begin{thebibliography}{99}

\footnotesize
\bibitem{review}
A. Pelissetto and E. Vicari,
cond-mat/0012164.

\bibitem{CFN-82}
J.-H.~Chen, M.E.~Fisher, and B.G.~Nickel,
Phys.\ Rev.\ Lett.\ {\bf 48} (1982) 630.

\bibitem{GR-84} M.J.~George and J.J.~Rehr,
Phys.\ Rev.\ Lett.\ {\bf 53} (1984) 2063.
 
\bibitem{FC-85}
M.E.~Fisher and J.H.~Chen,
J.\ Physique {\bf 46} (1985) 1645. 

\bibitem{BFMM-98}
H.G.~Ballesteros, L.A.~Fern\'andez, V.~Mart\'{\i}n-Mayor,
and A.~Mu\~noz Sudupe, Phys.\ Lett.\ B {\bf 441} (1998) 330.
 
\bibitem{HPV-99}
M.~Hasenbusch, K.~Pinn, and S.~Vinti,
Phys.\ Rev.\ B {\bf 59} (1999) 11471. 
 
\bibitem{BFMMPR-99}
H.G.~Ballesteros, L.A.~Fern\'andez, V.~Mart\'{\i}n-Mayor,
A.~Mu\~noz Sudupe, G.~Parisi, and J.J.~Ruiz-Lorenzo,
J.~Phys.\ A {\bf 32} (1999) 1.
 
\bibitem{Hasenbusch-99}
M.~Hasenbusch, J.~Phys.\ A {\bf 32} (1999) 4851.

\bibitem{HT-99}
M.~Hasenbusch and T.~T\"or\"ok,  J.~Phys.\ A {\bf 32} (1999) 6361.
 
\bibitem{CHPRV-01}
M.~Campostrini, M. Hasenbusch, A.~Pelissetto, P.~Rossi, and E.~Vicari,
Phys. Rev. B {\bf 63} (2001) 214503.
 
\bibitem{Hasenbusch-01}
M. Hasenbusch,
J. Phys. A {\bf 34} (2001) 8221. 

\bibitem{CHPRV-02}
M.~Campostrini, M. Hasenbusch, A.~Pelissetto, P.~Rossi, and E.~Vicari,
cond-mat/0110336.

\bibitem{CPRV-99}
M.~Campostrini, A.~Pelissetto, P.~Rossi, and E.~Vicari,
Phys.\ Rev.\ E {\bf 60} (1999) 3526; cond-mat/0201180.

\bibitem{CPRV-00}
M.~Campostrini, A.~Pelissetto, P.~Rossi, and E.~Vicari,
Phys.\ Rev.\ B {\bf 61} (2000) 5905.

\bibitem{BW-73} E.~Br\'ezin and D.J.~Wallace,
Phys.\ Rev.\ B {\bf 7} (1973) 1967.
 
\bibitem{BZ-76}
E.~Br\'ezin and J.~Zinn-Justin,
Phys. Rev. B {\bf 14} (1976) 3110.
 
\bibitem{WZ-75}
D.J.~Wallace and R.P.K.~Zia,
Phys.\ Rev.\ B {\bf 12} (1975) 5340.

\bibitem{SH-78}
L.~Sch\"afer and H.~Horner,
Z.~Phys.\ B {\bf 29} (1978) 251.
 
\bibitem{Lawrie-81}
I.D.~Lawrie, J.~Phys.\ A {\bf 14} (1981) 2489.

\bibitem{PV-99}
A.~Pelissetto and E.~Vicari, Nucl.\ Phys.\ B {\bf 540} (1999) 639.

\bibitem{Schofield-69}
P.~Schofield, Phys.\ Rev.\ Lett.\ {\bf 22} (1969) 606.
 
\bibitem{SLH-69}
P.~Schofield, J.D.~Lister, and J.T.~Ho,
Phys.\ Rev.\ Lett.\ {\bf 23} (1969) 1098.
 
\bibitem{Josephson-69}
B.D.~Josephson, J.~Phys.\ C: Solid State Phys.\ {\bf 2} (1969) 1113.

\bibitem{GZ-97}
R.~Guida and J.~Zinn-Justin,
Nucl.\ Phys.\ B {\bf  489} (1997) 626.

\bibitem{CHPV-01}
M. Caselle, M. Hasenbusch, A. Pelissetto, and E. Vicari,
J. Phys. A {\bf 34} (2001) 2923.

\bibitem{CPRV-00-2}
M.~Campostrini, A.~Pelissetto, P.~Rossi, and E.~Vicari,
Phys.\ Rev.\ B {\bf 62} (2000) 5843.


\bibitem{EHMS-00}
J.~Engels, S.~Holtmann, T.~Mendes, and T.~Schulze, 
Phys. Lett. B {\bf 492} (2000) 219.

\bibitem{LSNCI-96}
J.A. Lipa {\em et al.}, Phys.\ Rev.\ Lett.\ {\bf 76} (1996) 944;
{\bf 84} (2000) 4894.

\bibitem{HES-01}
S.~Holtmann, J.~Engels, and T.~Schulze,
hep-lat/0109013.

\bibitem{Nissen-etal-98}
J.A.~Nissen, D.R.~Swanson, Z.K.~Geng, V.~Dohm, U.E.~Israelsson,
M.J.~DiPirro, and J.A.~Lipa,
Low Temp.\ Phys.\ {\bf 24} (1998) 86.


\end{thebibliography}
\end{document}